\documentclass[aps,prd,twocolumn,10pt,superscriptaddress,nofootinbib,nobibnotes,longbibliography]{revtex4-1}
\usepackage{amssymb}
\usepackage{graphicx}
\usepackage{amsmath}
\usepackage{hyperref}
\usepackage{subfigure}
\usepackage{multirow}
\usepackage{setspace}
\usepackage{verbatim}
\usepackage{float}
\usepackage{color}
\usepackage{ulem}
\usepackage{bm}
\usepackage[utf8]{inputenc}

\begin{document}
	

\title{Reconciling early dark energy with a Harrison-Zeldovich spectrum}

\author{Chengjie Fu}
\email[]{fucj@ahnu.edu.cn}
\affiliation{ Department of Physics, Anhui Normal University, Wuhu, Anhui 241002, China }

\author{Shao-Jiang Wang} 
\email[Corresponding author:~]{schwang@itp.ac.cn}
\affiliation{ CAS Key Laboratory of Theoretical Physics, Institute of Theoretical Physics, Chinese Academy of Sciences, Beijing 100190, China }
\affiliation{Asia Pacific Center for Theoretical Physics (APCTP), Pohang 37673, Korea}

\begin{abstract}
Recent attempts to fully resolve the Hubble tension from early dark energy models seem to favor a primordial Harrison-Zeldovich universe with its scalar spectrum being extremely scale invariant. Restoring the Harrison-Zeldovich spectrum within the single-field inflationary paradigm appears to be infeasible, turning to the multifield approach from either curvaton or waterfall models. In this Letter, we successfully align with the Harrison-Zeldovich spectrum within a single-field chaotic inflation by a nonminimal derivative coupling, and the previously disfavored chaotic potential by Planck+BICEP/Keck data in the standard $\Lambda$-cold-dark-matter model now returns back to the scope of future polarization observations of the cosmic microwave background.
\end{abstract}

\maketitle

\section{Introduction}\label{sec:intro}

The standard $\Lambda$-cold-dark-matter ($\Lambda$CDM) might be cracked by the $5\sigma$ $H_0$ tension~\cite{Bernal:2016gxb,Verde:2019ivm,Knox:2019rjx,Riess:2020sih,DiValentino:2020zio,DiValentino:2021izs,Perivolaropoulos:2021jda,Abdalla:2022yfr}, $2\sim 3\sigma$ $S_8$ tension~\cite{DiValentino:2020vvd,Perivolaropoulos:2021jda,Abdalla:2022yfr}, and recently discovered $\sim 3\sigma$ $\delta H_0$ tension~\cite{Yu:2022wvg} and $3\sim7\sigma$ $a_B$ tension~\cite{Huang:2024erq}. Depending on how the source of discrepancy is recognized, the $H_0$ tension can be equivalently reformulated as the $r_s$ tension~\cite{Bernal:2016gxb,Verde:2019ivm,Knox:2019rjx,Riess:2020sih} or $M_B$ tension~\cite{Benevento:2020fev,Camarena:2021jlr,Efstathiou:2021ocp,Cai:2021weh,Cai:2022dkh} so as to modify the early or late Universe, respectively. However, various no-go arguments against the early-time~\cite{Jedamzik:2020zmd,Lin:2021sfs,Vagnozzi:2021gjh,Philcox:2022sgj} and late-time~\cite{Benevento:2020fev,Camarena:2021jlr,Efstathiou:2021ocp,Cai:2021weh,Cai:2022dkh} resolutions have led us possibly in the direction that either both early and late Universe should be modified altogether~\cite{Vagnozzi:2023nrq} or some new physics~\cite{Marra:2021fvf,Cai:2021wgv} hidden and disguised as ``systematics'' in the local Universe~\cite{Yu:2022wvg}, that is, before these ``systematics'' can be appropriately modeled, they are practically removed from the data by some empirical ansatz. On the other hand, the primordial Universe scenarios are less discussed as it appears too early to affect our current local expansion.

Intriguingly, almost all purely early resolutions by reducing the sound horizon $r_s$ alone without altering the post-recombination history tend to increase the Hubble constant $H_0$ at a price of increasing the scalar spectrum index $n_s$, $\delta n_s\simeq0.4(\delta H_0/H_0)\simeq-0.4(\delta r_s/r_s)$~\cite{Ye:2021nej,Jiang:2022uyg,Peng:2023bik}. For example, fitting the original early dark energy (EDE) model for an axionlike potential with the most preferable $n=3$ case to the full Planck+BAO+Pantheon+$H_0$ dataset gives rise to the mean (best-fit) value $\pm1\sigma$ error as $n_s=0.9812(0.9880)\pm0.0080$~\cite{Poulin:2018cxd}. Similarly, fitting the new EDE model~\cite{Niedermann:2019olb} to an almost the same dataset increases $n_s=0.9889(0.9912)_{-0.0066}^{+0.0067}$~\cite{Niedermann:2020dwg}. More extremal case comes from an early anti-de Sitter (AdS) phase of EDE that uplifts $n_s=0.9976(0.9974)_{-0.0045}^{+0.0046}$~\cite{Ye:2020btb}. It seems that a full reconciliation of the $5\sigma$ $H_0$ tension seems to prefer a primordial Harrison-Zeldovich universe with a scale-invariant scalar spectrum index $n_s=1$~\cite{Harrison:1969fb,Zeldovich:1972zz,Peebles:1970ag}.

However, the Planck data alone has already ruled out decisively the Harrison-Zeldovich spectrum at $6\sim8\sigma$ confidence level~\cite{Planck:2018jri} if the $\Lambda$CDM model is assumed throughout the cosmic history. What turns out unusually surprising is that both the axionlike EDE and AdS-EDE models with $n_s=1$ prior fit the Planck+BAO+Pantheon+$H_0$ dataset better than the $\Lambda$CDM model with the Bayes ratio $\Delta \ln B\simeq10$~\cite{Jiang:2022qlj}, while the absence of the local $H_0$ prior from the above dataset turns out the other way around with moderate evidence $\Delta \ln B\simeq-1\sim-3$ against the $n_s=1$ EDE scenario. This is different from other $H_0$-solution models adopting a local $H_0$ prior for data analysis simply to compromise between the global and local $H_0$ measurements without actually improving the Bayes ratio. Furthermore, adding other ground-based cosmic microwave background (CMB) polarization datasets would even enlarge the Bayes ratio to $\Delta \ln B\simeq15$ for the preference of $n_s=1$~\cite{Jiang:2022qlj}. This is not surprising as with these ground-based CMB polarization data alone, the $\Lambda$CDM model already favors $n_s=1$ over $n_s\sim0.965$ from Planck~\cite{Giare:2022rvg}.

If a primordial Harrison-Zeldovich Universe is indeed inferred from resolving the Hubble tension with EDE models, then the inflation model construction is thus desired for~\cite{Braglia:2020bym} as in the standard canonical single-field inflation, a scale-invariant spectral index $n_s\simeq 1-\mathcal{O}(1)/N_*\to1$ would require a much longer $e$-folding number than what we actually needed for solving the horizon problem. Therefore, the inflationary model building usually involves multifield configurations as proposed recently in, for example, axion curvaton~\cite{Takahashi:2021bti} and hybrid waterfall~\cite{Ye:2022efx,Braglia:2022phb} models (see also~\cite{Lin:2022gbl} for D-term inflation in braneworld scenario). Since no evidence has been reported yet for the multifield inflation, it would be more appealing to produce a primordial Harrison-Zeldovich universe within the single-field inflationary scenario.

In this Letter, we achieve this goal by turning on a nonminimal derivative coupling (NDC) introduced in Sec.~\ref{sec:NMC} with a typical example presented in Sec.~\ref{sec:chaotic}. We conclude in Sec.~\ref{sec:condis} with discussions in future perspectives.

\section{Nonminimal derivative coupling}\label{sec:NMC}

The NDC was originally introduced in the heterotic string theory for the universal dilaton~\cite{Gross:1986mw}. Later in Ref.~\cite{Germani:2010gm}, this NDC was borrowed phenomenologically as a unique coupling of the standard model Higgs boson to the Einstein gravity so as to yield a successful inflation within the standard model Higgs parameters without dangerous quantum corrections. In this Letter, we also approach the nonminimally derivative coupled scalar field phenomenologically but as a generic scalar field, leading to the ensuing action
\begin{align}\label{action}
 S=\int\mathrm{d}^4x \sqrt{-g}\left[\frac{M_{\rm Pl}^2}{2}R - \frac{1}{2}\left(g^{\mu\nu}- \xi G^{\mu\nu}\right)\nabla_\mu\phi\nabla_\nu\phi- V\right],
\end{align}
where $M_{\rm Pl}=1/\sqrt{8\pi G}$ is the reduced Planck mass, $G^{\mu\nu}$ is the Einstein tensor, and $\xi$ is a coupling constant with $\mathrm{mass}^{-2}$ dimension. Furthermore, the action \eqref{action} is part of a larger class of scalar-tensor theories~\cite{Kobayashi:2011nu,Deffayet:2011gz} that possess second-order equations of motion (EoM), propagating no more degrees of freedom than general relativity minimally coupled to a scalar field.
A salient characteristic of the NDC model is that with a positive coupling $\xi$, the inflaton evolution is slowed down compared to the minimally coupled case due to gravitationally enhanced friction, providing an avenue to reconcile a steep potential $V(\phi)$ with the CMB observations~\cite{Tsujikawa:2012mk}. For example, Refs.~\cite{Fu:2019ttf,Fu:2019vqc} managed to realize the ultra-slow-roll inflation that amplifies the small-scale primordial curvature perturbations by generalizing the coupling constant as a special function of the inflaton.

We first review the background dynamics~\cite{Tsujikawa:2012mk}.
In a spatially flat FLRW universe, characterized by $\mathrm{d}s^2=-\mathrm{d}t^2+a(t)^2\delta_{ij}\mathrm{d}x^i\mathrm{d}x^j$, the background dynamics for the NDC model is determined by the field equations, 
\begin{align}
 3M_{\rm Pl}^2H^2=\frac{1}{2}\left(1+9\xi H^2\right)\dot\phi^2+V(\phi), \label{HC}
\end{align}
\begin{align}
 \left(1+3\xi H^2\right)\ddot\phi + \left[1+\xi\left(2\dot H+3H^2\right)\right]3H\dot\phi+V_{,\phi}=0,\label{EoM}
\end{align}
where $V_{,\phi} = \mathrm{d}V/\mathrm{d}\phi$, and the overdot symbol denotes the derivative with respect to the cosmic time $t$. In the context of the slow-roll inflation, it is convenient to introduce the slow-roll parameters,
\begin{align}\label{SLP}
 \begin{aligned}
  \epsilon &= -\frac{\dot H}{H^2},\qquad &\delta_\phi&=\frac{\ddot\phi}{H\dot\phi}, \\
  \delta_X &=\frac{\dot\phi^2}{2M_{\rm Pl}^2H^2},&\delta_D&=\frac{\xi\dot\phi^2}{4M_{\rm Pl}^2},
 \end{aligned}
\end{align}
to characterize the background evolutions. During the slow-roll inflation, it is required that $\epsilon$, $|\delta_\phi|$, $\delta_X$, and $|\delta_D|\ll 1$. This allows us to simplify field equations as
\begin{align}
 &3M_{\rm Pl}^2H^2 \simeq V(\phi), \label{HC_approx}\\
 &3H \mathcal{A} \dot\phi+V_{,\phi} \simeq 0, \label{EoM_approx}
\end{align}
with an abbreviation
\begin{align}\label{def_A}
 \mathcal{A} = 1+3\xi H^2.
\end{align}
Taking the time derivative of Eq.~\eqref{HC_approx} and employing both Eqs.~\eqref{HC_approx} and~\eqref{EoM_approx}, we arrive at a relation
\begin{align}
 \epsilon\simeq \delta_X + 6\delta_D \simeq \frac{\epsilon_V}{\mathcal{A}}
\end{align}
to the usual first potential slow-roll parameter
\begin{align}
\epsilon_V = \frac{M_{\rm Pl}^2}{2}\left( \frac{V_{,\phi}}{V} \right)^2.
\end{align}
The field value $\phi_e$ that terminates the slow-roll inflation is ascertained by fulfilling the condition $\epsilon(\phi_e)=1$, namely $\epsilon_V(\phi_e)\left[ 1 + \xi M_{\rm Pl}^{-2} V(\phi_e)\right]^{-1} = 1$. Hence, the \textit{e}-folding number $N$ from some moment during slow-roll inflation to the end of inflation can be approximated as
\begin{align}\label{eq:Nefold}
N(\phi) = \int^{\phi_e}_{\phi}\frac{H}{\dot{\tilde{\phi}}}d\tilde\phi \simeq \frac{1}{M_{\rm Pl}^2} \int^{\phi}_{\phi_e} \left(1 + \xi M_{\rm Pl}^{-2} V(\tilde{\phi}) \right) \frac{V(\tilde{\phi})}{V_{,{\tilde\phi}}} d\tilde\phi.
\end{align}

We next collect the perturbative dynamics~\cite{Tsujikawa:2012mk} for the scalar perturbations. The curvature perturbations $\mathcal{R}$ in the momentum space follows the EoM as
\begin{align}\label{R_k}
	 \mathcal{\ddot R}_k + \left( 3H + \frac{\dot{Q}_s}{Q_s} \right) \mathcal{\dot R}_k + \frac{c_s^2k^2}{a^2}\mathcal{R}_k = 0,
\end{align}
where the abbreviations
\begin{align}\label{Qs}
	Q_s=\frac{w_1(4w_1w_3+9w_2^2)}{3w_2^2},
\end{align}
\begin{align}\label{Cs2}
	c_s^2=\frac{3(2w_1^2w_2H-w_2^2w_4+4w_1\dot w_1w_2-2w_1^2\dot w_2)}{w_1(4w_1w_3+9w_2^2)},
\end{align}
are defined from
\begin{align}\label{w}
 w_1&=M_\mathrm{Pl}^2(1-2\delta_D),\nonumber \\
 w_2&=2H M_\mathrm{Pl}^2(1-6\delta_D),\nonumber \\
 w_3&=-3H^2M_\mathrm{Pl}^2(3-\delta_X-36\delta_D), \\
 w_4&=M_\mathrm{Pl}^2(1+2\delta_D).\nonumber
\end{align}
Under the slow-roll approximation, one can deduce that $Q_s\simeq M_\mathrm{Pl}^2 \epsilon_V/\mathcal{A}$ and $c_s^2 = 1 - \mathcal{O}(\epsilon)$.
The power spectrum and the spectral index for the curvature perturbations, evaluated at $c_sk=aH$, can be estimated as
\begin{align}
\mathcal{P}_\mathcal{R} = \frac{H^2}{8\pi^2Q_s c_s^3} \simeq \frac{V^3}{12\pi^2M_\mathrm{Pl}^6V_{,\phi}^2} \left(1 + \xi M_{\rm Pl}^{-2} V\right) ,
\end{align} 
\begin{align}\label{n_s}
n_s-1=\frac{d\ln \mathcal{P_R}}{d\ln k}\simeq-\frac{1}{\mathcal{A}}\left[2\epsilon_V\left(4-\frac{1}{\mathcal{A}}\right)-2\eta_V\right],
\end{align} 
respectively, with the use of the second potential slow-roll parameter 
$\eta_V = M_\mathrm{Pl}^2 V_{,\phi\phi}/V$.

We finally turn to the perturbative dynamics~\cite{Tsujikawa:2012mk} for the tensor perturbations $h_{ij}$. After expanded in terms of two independent transverse traceless basis tensors,
\begin{align}
h_{ij}(t,\boldsymbol{x})=\sum\limits_{A=\pm} \int \frac{d^3k}{(2\pi)^{3/2}} h^A_k(t)e^A_{ij}(k)e^{i\boldsymbol{k}\cdot\boldsymbol{x}},
\end{align}
with $e^A_{ij}(k)e^{A^\prime}_{ij}(k)=2\delta_{AA^\prime}$, its Fourier modes $h^A_k$ obey
\begin{align}\label{h_k}
	{\ddot h}^A_k + \left( 3H + \frac{\dot{Q}_t}{Q_t} \right) {\dot h}^A_k + \frac{c_t^2k^2}{a^2}{h}^A_k = 0,
\end{align}
with
\begin{align}
Q_t &= w_1/4 = M_\mathrm{Pl}^2(1-2\delta_D)/4,\\
c_t^2 &= w_4/w_1 = 1 + 4\delta_D + \mathcal{O}(\epsilon^2).
\end{align}
The tensor power spectrum evaluated at $c_tk=aH$ reads
\begin{align}
\mathcal{P}_T = \frac{H^2}{2\pi^2Q_tc_t^3} \simeq \frac{2V}{3\pi^2M_\mathrm{Pl}^4}.
\end{align}
Therefore, the tensor-to-scalar ratio can be estimated as
\begin{align}
r=\left. \frac{\mathcal{P_R}}{\mathcal{P}_T} \right|_{k\simeq aH} \simeq \frac{16\epsilon_V}{\mathcal{A}}.
\end{align}

\begin{figure}
	\centering
	\includegraphics[width=0.9\columnwidth ]{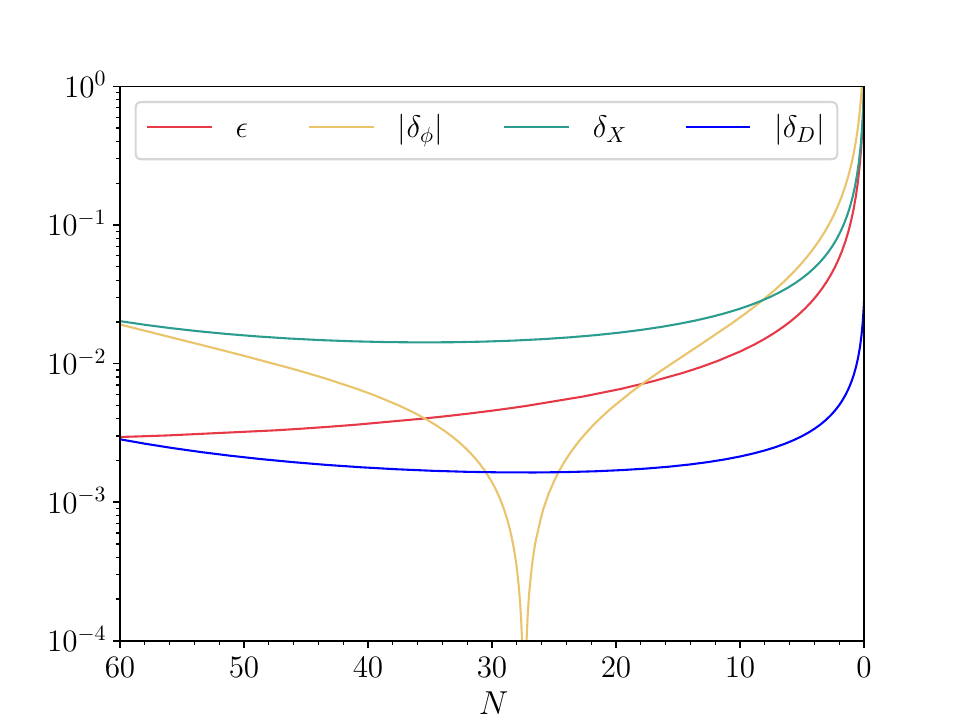}
	\caption{\label{fig1}  The evolution of the slow-roll parameters defined in Eq. \eqref{SLP} as a function of the \textit{e}-folding number $N$.}
\end{figure}

\section{Chaotic potential with negative nonminimal derivative coupling}\label{sec:chaotic}

In this section, we exemplify a realization of primordial Harrison-Zeldovich universe in the NDC model for the simplest inflationary potential, that is, the chaotic potential with a monomial power-law form,
\begin{align}
	V(\phi) = \lambda M_{\rm Pl}^{4-p} \phi^p,
\end{align}
where the dimensionless parameter $\lambda$ sets the inflationary energy scale solely determined by the amplitude of the primordial scalar spectrum at the CMB scale. The inflationary dynamics is fixed once the rescaled coupling constant $\gamma \equiv \lambda\xi M_{\rm Pl}^2$ and the power $p$ are specified. 

The latest CMB results from the combined observations of Planck 2018 \cite{Planck:2018jri} and BICEP/Keck \cite{BICEP:2021xfz} assuming the standard $\Lambda$CDM model has unequivocally ruled out the entire family of monomial power-law potentials within the minimally coupled canonical framework~\cite{Mishra:2022ijb}. Notably, this conclusion remains steadfast even for the NDC model with a positive coupling $\xi$~\cite{Avdeev:2022ilo}. The negative coupling $\xi$ is less considered before so as to avoid the ghost propagation. What turns out as a nice surprise is that, alongside with a primordial Harrison-Zeldovich universe, the chaotic potential in the NDC model with a negative coupling $\xi$ is not as dangerous as it appears to be, since the ghost propagation for the curvature perturbations never occur and the negative kinetic term would eventually evolve into a canonical form before the end of the inflation as shown shortly below.

Let us first locate the parameter region allowed for a primordial Harrison-Zeldovich universe. Incorporating $\epsilon_V = (p^2/2)(\phi/M_{\rm Pl})^{-2}$ and $\eta_V = p(p-1)(\phi/M_{\rm Pl})^{-2}$ into Eq. \eqref{n_s} renders the scalar spectral index as
\begin{align}
 n_s\simeq 1 - \frac{p}{\mathcal{A}}\left( 2p - \frac{p}{\mathcal{A}} + 2 \right) \left(\frac{\phi}{M_{\rm Pl}}\right)^{-2},
\end{align}
which can be extremely scale invariant $n_s=1$ at the CMB scale as long as the condition $( 2p - p/\mathcal{A}_\ast + 2 ) =0$ is imposed. Here the quantity with subscript $\ast$ is specifically evaluated at $\phi=\phi_\ast$ when the CMB scale exits the horizon. Hence, we require
\begin{align}\label{A_ast}
\mathcal{A}_\ast = \frac{p}{2(p+1)}.
\end{align} 
It is easy to see that $0<\mathcal{A}_\ast < 1 $ holds true for any positive potential power $p>0$, thus $\xi$ must be negative to ensure $\mathcal{A}_\ast=1+3\xi H_*^2<1$. Then, as our Universe inflates with a slowly decreasing $H$, a negative $\xi$ would further enlarge $\mathcal{A}=1+3\xi H^2$ from a positive $\mathcal{A}_\ast>0$ so that $\mathcal{A}>0$ namely $Q_s\simeq M_\mathrm{Pl}^2\epsilon_V/\mathcal{A}>0$ can be always guaranteed during inflation, effectively precluding the emergence of the ghost propagation for the curvature perturbations.

We next carry out the inflationary predictions for our chaotic potential with a negative NDC coupling within the scale-invariant parameter space specified by the condition~\eqref{A_ast}. The $e$-folding number $N_*$ from Eq.~\eqref{eq:Nefold} can be related to the field value $\phi_\ast$ by
\begin{align}\label{N_ast}
N_\ast &\simeq \frac{p+2\mathcal{A}}{2p(p+2)}\left( \frac{\phi}{M_{\rm Pl}} \right)^2 \bigg|^{\phi_\ast}_{\phi_e} \simeq \frac{p+2\mathcal{A}_\ast}{2p(p+2)}\left( \frac{\phi_\ast}{M_{\rm Pl}} \right)^2 \nonumber \\
& \simeq \frac{1}{2(p+1)} \left( \frac{\phi_\ast}{M_{\rm Pl}} \right)^2,
\end{align}
where we have used Eqs.~\eqref{HC_approx}, \eqref{def_A}, and~\eqref{A_ast} as well as the fact $\phi_\ast \gg \phi_e$. Combining~\eqref{A_ast} with~\eqref{N_ast}, the tensor-to-scalar ratio at the CMB scale can be estimated as
\begin{align}\label{r_ast}
	r = \frac{8p}{N_\ast}.
 \end{align}
Drawing from Eqs.~\eqref{HC_approx}, \eqref{def_A}, \eqref{A_ast}, and~\eqref{N_ast}, we can further estimate the rescaled coupling constant $\gamma$ as
\begin{align}\label{gamma}
\gamma = - \frac{p+2}{[2(p+1)]^{p/2+1}}\frac{1}{N_\ast^{p/2}}.
\end{align}

\begin{figure}
	\centering
	\includegraphics[width=0.9\columnwidth ]{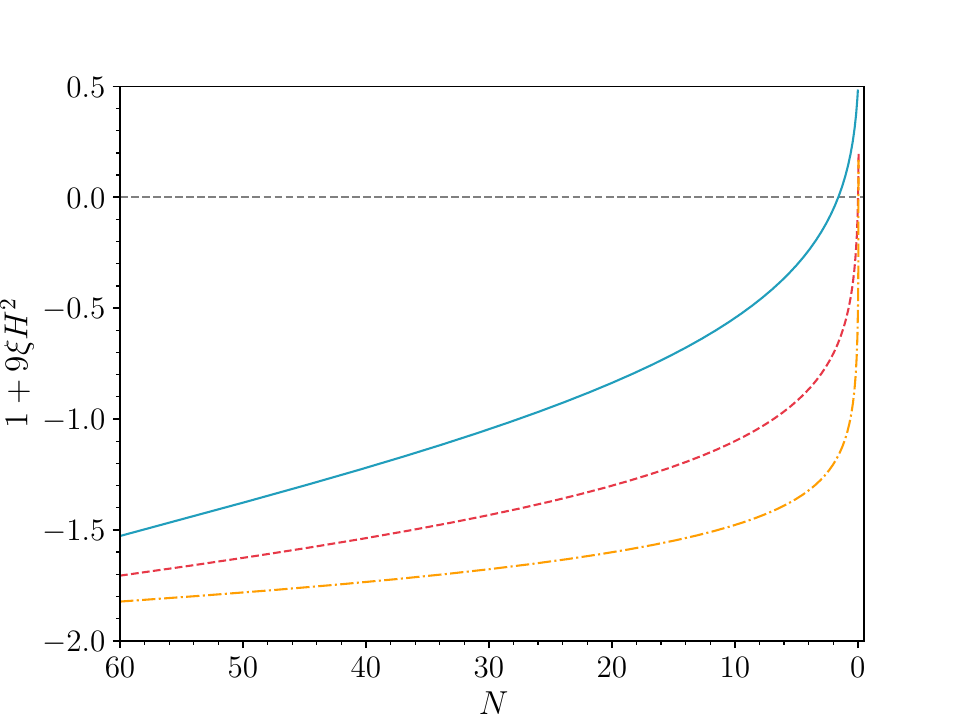}
	\caption{\label{fig2}  The evolution of the correction factor in the kinetic term, $(1+9\xi H^2)$, as a function of the $e$-folding number $N$ in the cases of $p=2/5$ (solid line), $p=1/5$ (dashed line), and $p=1/10$ (dash-dot line). }
\end{figure}

Finally, we arrive at presenting our numerical results based on solving the intertwined background Eqs.~\eqref{HC} and~\eqref{EoM} alongside the perturbation Eqs.~\eqref{R_k} and~\eqref{h_k}. Taking $p=2/5$ for a simple illustration, the theoretical computation~\eqref{gamma} for the rescaled coupling constant yields $\gamma = - 0.308$ from setting $N_\ast = 60$, which also fixes $\phi_\ast = 12.962 M_{\rm Pl}$ by Eq.~\eqref{N_ast}, aligning closely with the exact numerical determination $\phi_\ast = 12.697 M_{\rm Pl}$. This concurrence can be attributed to the fact that the slow-roll conditions $\{ \epsilon,|\delta_\phi|, \delta_X, |\delta_D|\} \ll 1$ are perfectly maintained throughout an inflationary duration of $60$ $e$-folds as verified numerically in Fig.~\ref{fig1} for the time evolution of these slow-roll parameters. 

Another potential concern is the kinetic term of the background field equation~\eqref{HC}, whose prefactor $(1+9\xi H^2)$ emerges as negative at the onset of inflation. This wrong-sign kinetic term would typically lead to ghost instability if such a regime persists. Fortunately, for the example explored herein, we have explicitly checked with exact numerical solutions for the prefactor $(1+9\xi H^2)$ evolution as illustrated in Fig.~\ref{fig2} for $p=2/5$, which always manifests a transition from a negative value to a positive one just prior to the end of inflation, avoiding ghost instability and guaranteeing a graceful exit from inflation. 
It is intriguing to observe that for adequately small values of $p$, this negative-to-positive transition of the prefactor $(1+9\xi H^2)$ nearly coincides with the end of inflation, as illustrated in Fig.~\ref{fig2}, which also includes the cases for $p=1/5$ with $\gamma=-0.558$ and $p=1/10$ with $\gamma=-0.748$.

Before confronting our model with CMB observations, we have to further check whether our analytical analysis for achieving $n_s=1$ is reliable. It turns out that the exact numerical calculations for the example prementioned give rise to $n_s=0.9958$ and $r=0.051$ at the CMB scale, which are nicely consistent with their theoretical counterparts $n_s=1$ and $r=0.053$ from our analytic estimation~\eqref{r_ast}. This ensures that with appropriately fine-tuning the negative NDC, we can always arrive at a Harrison-Zeldovich spectrum in our NDC model.

\begin{figure*}
\centering
\includegraphics[width=1.\textwidth ]{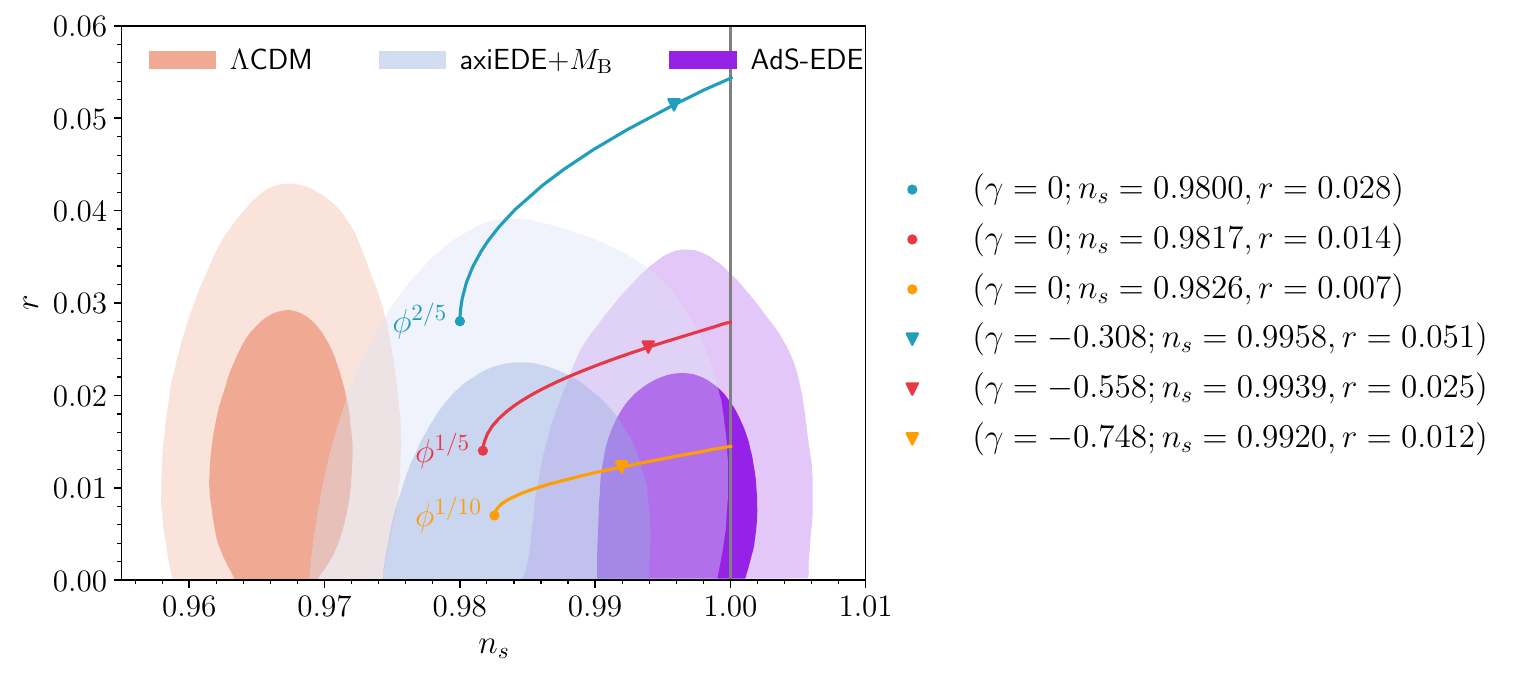}
\caption{\label{fig3}  The $n_s-r$ predictions for our NDC model with a negative NDC coupling $\xi$ and a power-law chaotic potential $V(\phi)=\lambda M_\mathrm{Pl}^{4-p}\phi^p$. The effective coupling $\gamma\equiv\lambda\xi M_\mathrm{Pl}^2$ varies between  $-0.3114\leqslant\gamma\leqslant0$, $-0.5652\leqslant\gamma\leqslant0$, and $-0.75798\leqslant\gamma\leqslant0$ for $p=2/5, 1/5, 1/10$, respectively, with the lower bound of $\gamma$ corresponding to an exact scale-invariant case $n_s=1$ as shown with a gray solid line. The triangle points are associated with the adopted values of $\gamma$ obtained from our analytic estimation~\eqref{gamma}, while the dotted points are the corresponding minimal coupling case with $\gamma=0$. The shaded regions depict 68\% and 95\% C.L. contours of $n_s-r$ for the $\Lambda$CDM model (orange shaded)~\cite{BICEP:2021xfz}, the AdS-EDE model (purple shaded)~\cite{Ye:2022afu}, and the axionlike EDE (axiEDE) model (blue shaded)~\cite{Camarena:2021jlr,Efstathiou:2021ocp} with an equivalent $H_0$ prior from the SH0ES calibration on the absolute magnitude $M_{\rm B}$ of type Ia supernovae.}
\end{figure*}

At last in Fig.~\ref{fig3}, we can compare with the CMB observation in the $n_s-r$ plane as a function of the effective coupling $\gamma$ involving a negative NDC $\xi$ for our NDC model with a chaotic potential for three illustrative powers $p=2/5, 1/5, 1/10$, where the corresponding effective couplings range within $-0.3114\leqslant\gamma\leqslant0$, $-0.5652\leqslant\gamma\leqslant0$, and $-0.75798\leqslant\gamma\leqslant0$, respectively, interpolating between the Harrison-Zeldovich spectrum and minimally coupled cases. The orange, purple, and blue shaded regions are the cosmological constraints~\cite{Ye:2022afu} from the dataset Planck 2018 $+$ BICEP/Keck $+$ BAO $+$ Pantheon for the $\Lambda$CDM model~\cite{BICEP:2021xfz}, AdS-EDE model~\cite{Ye:2022afu}, and axionlike EDE model (with an equivalent $H_0$ prior)~\cite{Camarena:2021jlr,Efstathiou:2021ocp}, respectively. It is evident from Fig.~\ref{fig3} that, with decreasing the effective coupling $\gamma$, the scalar spectral index becomes close to the scale-invariant case $n_s=1$ but at a price of enlarging the tensor-to-scalar ratio $r$, which has been strongly constrained by the current CMB polarization observations. However, for a chaotic potential with shallower power $0<p\ll1$, one can always achieve a Harrison-Zeldovich spectrum and at the same time meet the upper bound on the tensor-to-scalar ratio.

\section{Conclusions and discussions}\label{sec:condis}

The EDE resolutions to the Hubble tension necessarily tend to increase the scalar spectral index extremely close to the scale-invariant case. In particular, when a Harrison-Zeldovich spectrum is assumed for these EDE models, they become even more preferred by the dataset CMB+BAO+SNe+$H_0$ with higher Bayes evidence than the standard $\Lambda$CDM does, but without the local $H_0$ prior, these EDE models show no preference over the standard $\Lambda$CDM. This intriguing phenomenon revives inflationary model buildings with a Harrison-Zeldovich spectrum for these EDE models but usually within the multifield scenario. In this Letter, we propose a single-field inflationary model with a negative NDC for a chaotic power-law potential, $V(\phi) \propto \phi^p$, that has already been ruled out in the standard $\Lambda$CDM. Our illustrative model for a shallower potential with $0<p\ll1$ has brought back these chaotic potentials to the future scope from CMB polarization observations~\cite{Jiang:2023bsz}. It is worth noting that these fractional power-law potential with almost arbitrary rational values for the power $p$ can be dynamically generated in the framework of dynamical chaotic inflation from the dynamics of a strongly coupled supersymmetric gauge theory~\cite{Harigaya:2012pg,Harigaya:2014sua,Harigaya:2014wta}. Nevertheless, despite the success of landing EDE on a primordial Harrison-Zeldovich universe, the EDE models as like any other early resolutions to the Hubble tension would necessarily worsen the $S_8$ tension, thus calling for late Universe modifications at the same time to reduce the $S_8$ tension separately. A return of a Harrison-Zeldovich spectrum would also affects the late-time matter power spectrum at smaller scales, possibly leading to alternative observational constraints for future investigations.

\begin{acknowledgments}
We thank Zu-Cheng Chen and Lang Liu for fruitful discussions.
C.J. F. is supported by the National Key Research and Development Program of China Grant No. 2020YFC2201502, and the National Natural Science Foundation of China Grant No. 12305057.
S.J. W. is supported by the National Key Research and Development Program of China Grants No. 2021YFC2203004, No. 2020YFC2201501, and No. 2021YFA0718304, 
the National Natural Science Foundation of China Grants 
No. 12105344, No. 12235019, and No. 12047503,
the Key Research Program of the Chinese Academy of Sciences (CAS) Grant No. XDPB15, 
the Key Research Program of Frontier Sciences of CAS, 
and the Science Research Grants from the China Manned Space Project with Grant No. CMS-CSST-2021-B01.
\end{acknowledgments}



\bibliography{ref}

\end{document}